\documentstyle[twocolumn,aps,epsfig]{revtex}

\begin{document}

\title{Trapping Black Hole Remnants}
 
\author{Sabine Hossenfelder${}^1$\thanks{sabine@physics.arizona.edu}, 
Benjamin Koch${}^{2,3}$\thanks{koch@th.physik.uni-frankfurt.de} and 
Marcus Bleicher${}^2$\thanks{bleicher@th.physik.uni-frankfurt.de}, }

\address{${}^1$ Department of Physics\\ 
University of Arizona\\
1118 East 4th Street\\ 
Tucson, AZ 85721, USA}

\address{\vspace*{.3cm}${}^2$ Institut f\"ur Theoretische Physik\\ 
J. W. Goethe-Universit\"at\\
Max von Laue Strasse 1\\ 
60438 Frankfurt am Main, Germany}

\address{${}^3$ Frankfurt International Graduate School for Science (FIGSS)\\ 
J. W. Goethe-Universit\"at\\
Max von Laue Strasse 1\\ 
60438 Frankfurt am Main, Germany}
 
\maketitle

\noindent
\begin{abstract}
Large extra dimensions lower the Planck scale to values soon accessible. 
The production of TeV mass black holes at 
the {\sc LHC} is one of the most exciting predictions. However, the final 
phases of the black hole's evaporation are still unknown and there are 
strong indications that a black hole remnant can be left. Since a certain 
fraction of such objects would be electrically charged, we argue that they 
can be trapped. In this paper, we examine the occurrence of such charged 
black hole remnants.

These trapped remnants are of high interest, as they could be used to 
closely investigate the evaporation characteristics.
Due to the absence of background from the collision region and the controlled initial
state, the signal would
be very clear. This would allow to extract information about the late stages of the
evaporation process with high precision.
\end{abstract}

\section{Extra Dimensions}
Models with large extra dimensions ({\sc LXD}s)   
are motivated by string theory \cite{Antoniadis:1990ew} and provide us 
with an effective description of physics beyond the standard model (SM) 
in which observables can be computed and predictions can be made. 
Arkani-Hamed, Dimopoulos and Dvali \cite{Arkani-Hamed:1998rs} proposed 
a solution to the
hierarchy problem by the introduction of $d$ 
additional compactified space-like dimensions 
in which only the gravitons can propagate. The SM particles 
are bound to our 4-dimensional
sub-manifold, often called our 3-brane. 

This yields an attractive and simple explanation
of the hierarchy problem. Consider a particle of mass $M$ located in a space time with $d+3$ dimensions.
The general solution of Poisson's equation yields its potential as a function of the radial distance $r$ to the
source
\begin{eqnarray}
\phi(r) \propto \frac{1}{M_{\rm f}^{d+2}} \frac{M}{r^{d+1}} \quad, \label{newtond}
\end{eqnarray}
where a new fundamental mass-scale,  
$M_{\rm f}$, has been introduced.  
    
The additional $d$ space-time dimensions are compactified on radii $R$, which are
small enough to have been unobserved so far. Then, at distances $r \gg R$   the potential Eq. (\ref{newtond})
will turn into the common $1/r$ potential, but with a pre-factor given by the volume of the extra dimensions
\begin{eqnarray}
\phi(r) \to \frac{1}{M_{\rm f}^{d+2}} \frac{1}{R^{d}} \frac{M}{r} \quad. \label{newton} 
\end{eqnarray}
In the limit of large distances,  rediscovering the usual gravitational law yields the relation
\begin{eqnarray}
m_{\rm p}^2 = M_{\rm f}^{d+2} R^d \quad. \label{Master}
\end{eqnarray}
It can be seen from this argument that the volume of the extra dimensions suppresses the 
fundamental scale and thus, can explain the huge value of the observed Planck mass $m_{\rm p} \sim 10^{16}$~TeV.

The radius $R$ of these extra dimensions, for $M_{\rm f}\approx$~TeV, can be estimated with Eq.(\ref{Master}) and
typically lies in the range from $10^{-1}$~mm to $10^3$~fm for $d$ from $2$ to $7$, or the inverse radius 
$1/R$ lies in energy range eV to MeV, respectively.  
 
{\sc LXDs} predict a vast number of new effects at energies close to
the new fundamental scale. Among the signatures are modifications of SM observables due
to virtual graviton exchange, the production of real gravitons and TeV-scale black holes. 
For recent constraints from collider searches and astrophysics see e.g. \cite{Cheung:2004ab}.

\section{Black Holes in Extra Dimensions}

Using the higher dimensional Schwarzschild-metric \cite{my}, it can be derived that the
horizon radius $R_H$ of a black hole is substantially increased in the presence of
{\sc LXD}s, reflecting the fact
that gravity at small distances becomes stronger. For a black hole of mass $M$ one finds
\begin{eqnarray} \label{ssr}
R^{d+1}_H= \frac{2}{d+1} \frac{M}{M_{\rm f}^{d+2}} \quad.
\end{eqnarray}
The horizon radius for a black hole with mass $\approx$~TeV is then $\approx 10^{-3}$~fm, and thus
$R_H \ll R$ for black holes which might possibly be produced at colliders or in ultra high energetic 
cosmic rays.

Black holes with masses in the range of the lowered Planck scale should be a subject of
quantum gravity. Since there is yet no theory available to perform this analysis, 
the black holes are treated as semi classical objects. The gravitational field is
classical, though the evaporation process is a quantum effect.

To compute the production details, the cross-section of the black holes can be approximated
by the classical geometric cross-section 
\begin{eqnarray} \label{cross}
\sigma(M)\approx \pi R_H^2 \quad,
\end{eqnarray}
an expression which contains only  the fundamental Planck scale as coupling constant. This  
cross section is a subject of ongoing research \cite{Voloshin:2001fe} but close
investigations justify the use of the classical limit at least up to energies of 
$\approx 10 M_{\rm f}$ \cite{Solodukhin:2002ui}. It has further been shown that the naively 
expected classical result remains valid also in string theory 
\cite{Polchi}. 

A common approach to improve the naive picture of colliding point particles,
is to treat the creation of the horizon as a 
collision of two shock fronts in an Aichelburg-Sexl geometry describing the fast moving particles
 \cite{Rychkov:2004sf,GrWav}. Due to the high velocity of the moving particles, space time
before and after the shocks is almost flat and the geometry 
can be examined for the occurrence of trapped surfaces.

These semi classical considerations do also give rise to
form factors which take into account that not the whole initial energy is
captured behind the horizon. These factors have been calculated in \cite{Formfactors}, 
depend on the number of extra dimensions, and are of order one. They are included in our
numerical evaluation. The scattering of the emitted particle on the gravitational
potential is also taken into account in the graybody factors \cite{Solo}. 
 
Setting $M_{\rm f}\sim 1$TeV and $d=2$ one finds 
$\sigma \approx 1$~TeV$^{-2}\approx 400$~pb.
With this it is further found that these black holes 
will be produced at {\sc LHC}. For the estimated luminosity of 100/fb more than $\approx 10^8$ 
black holes are expected per year \cite{dim}. 
The potential importance of the black hole production at colliders has been 
investigated in numerous publications, see e.g. \cite{bhsonst,Chamblin:2002ad} 
and References therein.

Once produced, the black holes will undergo an evaporation process whose thermal
properties carry information about the parameters $M_{\rm f}$ and $d$. An analysis of the evaporation
will therefore offer the possibility to extract knowledge about the topology of our
space time and the underlying theory.

The evaporation process can be categorised in three characteristic stages \cite{Giddings3}:
the balding phase, the evaporation phase and the Planck phase. It is generally
assumed that in the last phase, the black hole will either completely decay in a few 
SM particles or a stable remnant will be left, which carries away the remaining energy. 
In the following we will focus on the case in which a remnant of approximately Planck mass is left.
 
For more details on the topic of TeV-scale black holes, 
the interested reader is referred to the reviews \cite{Kanti:2004nr}.

\section{Black Hole Remnants}

The final fate of these evaporating black holes is closely connected to the  
information loss puzzle. The black hole emits thermal radiation, whose sole 
property is the temperature, regardless of the initial state of the collapsing matter. 
So, if the black hole completely decays into statistically
distributed particles, unitarity can be violated. This happens when 
the initial state is a pure quantum state and then evolves into a mixed 
state \cite{Hawk82}.

When one tries to avoid the information loss problem two possibilities are left. 
The information
is regained by some unknown mechanism or a stable black hole remnant is formed 
which keeps the
information. Besides the fact that it is unclear in which way the information should escape the horizon 
\cite{escape} there are several other arguments for 
black hole remnants \cite{relics}:
\begin{itemize}
\item
The uncertainty relation: The Schwarzschild radius of a black hole with Planck mass 
is of the order  of the Planck length. Since the Planck length is the wavelength corresponding to a particle of
Planck mass, a problem arises when the mass of the black hole drops below Planck mass. 
Then one has trapped a mass inside a volume which is smaller than allowed by the uncertainty 
principle \cite{39}. To avoid this problem, Zel'dovich has proposed that black holes with masses below 
Planck mass should be associated with stable elementary particles \cite{40}. 

\item
Corrections to the Lagrangian: The introduction of additional terms, which are quadratic in the curvature, 
yields a dropping of the evaporation temperature towards zero \cite{Barrow}. This holds also for 
extra dimensional scenarios \cite{my2} and is supported  by calculations in the low energy limit 
of string theory \cite{Callan}.The production of TeV-scale 
black holes in the presence of Lovelock higher-curvature terms has been examined 
in \cite{Rizzo:2005fz} and it was found that these black holes can become 
thermodynamically stable since their evaporation takes an infinite amount of time.

\item
Further reasons for the existence of remnants have been suggested: E.g. black holes with axionic charge
\cite{axionic}, the modification of the Hawking temperature due to quantum hair \cite{hair} or magnetic 
monopoles \cite{magn}. Coupling of a dilaton field to gravity also yields remnants, with detailed 
features depending on the dimension of space-time \cite{dilaton1}.
 
\end{itemize}

Of course these remnants, which have also been termed Maximons, Friedmons, Cornucopions, Planckons or 
Informons, 
are not a miraculous remedy but bring some problems on their own, e.g.
the necessity for an infinite number of states which allows the unbounded information
content inherited from the initial state.

In \cite{first}, it has been shown that the black hole remnants give rise to 
distinct collider signatures.
 In the present work we will in particular
examine the
properties of  charged remnants.

\section{Charged Black Holes}

The black hole produced in a proton-proton collision can carry 
an electric charge. The evaporation spectrum contains all particles of the SM and so,
a certain fraction of the final black hole remnants will also carry net electric charge. In the following, these 
charged black hole remnants will be denoted $BH^+$ and $BH^-$, and the neutral ones $BH^0$, respectively.
Since the $BH^\pm$'s undergo an electromagnetic interaction, their cross section is enhanced and
they can be examined closely. This makes them extremely interesting candidates for the investigation of
Planck scale physics. 

The metric of a charged black hole in higher dimensions has been derived in \cite{my}. This
solution assumes the electric field to 
be spherical symmetric in all dimensions whereas in the scenario with {\sc LXD}s the
SM fields are confined to our brane. This has also been pointed out in Ref. \cite{Casadio:2001wh}. 

The exact solution for this system in a spacetime with compacitfied extra dimensions is known only implicitely \cite{higherdbhs}. However, for our purposes, it will be sufficient to estimate the charge effects 
by taking into account that we expect the brane to have a width 
of about $1/M_{\rm f}$. Up to this width, also the gauge fields can penetrate the bulk which is
essentially a scenario of embedding universal extra dimensions as a fat brane into the 
large extra dimensions.  
Note, that this does not modify the electromagnetic coupling constants as there is no
hierarchy between the inverse width and the radius of the fat brane. Following the same
arguments leading to the Newtonian potential (Eq.(\ref{newtond})), we see that the 
Coulomb potential receives a modification to
\begin{eqnarray} \label{higherc}
\phi_{\rm C} =   \frac{\alpha}{M_{\rm f}^{d+1}}\frac{Q}{r^{d+1}} \quad,
\end{eqnarray}
where $\alpha$ is the fine 
structure constant and $Q$ is the dimensionless charge in units of the unit charge $e$. But this
higher dimensional potential will already turn into the usual $1/r$ potential at a distance $r=1/M_{\rm f}$ which
means that the pre-factors cancel and $\alpha$ does not collect any volume factors.

One can estimate the exact solution for the system by assuming it to be spherical symmetric
up to the horizon radius. This yields
\begin{eqnarray}
g_{tt} = \gamma(r) =   1 - 2 \phi(r) \quad,
\end{eqnarray}
where $\phi$ is the potential containing the gravitational energy and the Coulomb energy
of the source whose electric field is now also higher dimensional. The weak field limit of
Einsteins field equations yields the Poisson equation
\begin{eqnarray}
- \Delta \phi(r) =   \Omega_{(d+3)} \delta(r) \frac{M}{M_{\rm f}^{d+2}} 
+ \frac{\alpha}{M_{\rm f}^{2(d+1)}} \frac{Q^2}{r^{2(d+2)}}  \quad,
\end{eqnarray}
where $\Omega_{(d+3)}$ is the surface of the $d+3$-dimensional unit sphere
\begin{eqnarray}
\Omega_{(d+3)} = \frac{2 \pi^{\frac{d+3}{2}}}{\Gamma({\frac{d+3}{2}})}\quad,
\end{eqnarray}
and the delta-function is already converted into spherical coordinates.
Using the spherical symmetry and applying Gauss' Law yields then
\begin{eqnarray}
\partial_r \phi(r) =    - \frac{M}{M_{\rm f}^{d+2}} \frac{1}{r^{d+2}} 
+ \frac{\alpha}{M_{\rm f}^{2(d+1)}} \frac{Q^2}{d+1} \frac{1}{r^{2d+3}}\quad.
\end{eqnarray}
And so we find
\begin{eqnarray} \label{phi}
\phi(r) =  \frac{1}{d+1} \frac{1}{M_{\rm f}^{d+2}} \frac{1}{r^{d+1}} 
\left[ 
M 
- 
\frac{\alpha Q^2}{2(d+1)} 
\frac{M_{\rm f}^{-d}}{ r^{d+1}} 
\right] \quad. 
\end{eqnarray}

The horizon $R_H$ is located at the zero
of $\gamma(r)$. In case there exists no (real) solution for $R_H$, the metric is dominated
by the contribution of the electromagnetic field and the singularity will be 
a naked one.  
The requirement of $\gamma$ having a zero yields the constraint\footnote{Note, that these relations do
only agree with the relations in \cite{my} up to geometrical pre-factors. This is due to the fact that
our additional dimensions are compactified and the higher dimensional coupling constants are fixed
by Eq (\ref{Master}) and (\ref{higherc}).}
\begin{eqnarray}
\alpha Q^2 \leq \left(\frac{M}{M_{\rm f}}\right)^{2} \quad.
\end{eqnarray}

With $M=$~few $\times M_{\rm f}$, and $Q$ being close by $e$, the left hand side is at least by a factor $100$ smaller than
the right hand
side.
So, the charge contribution to the gravitational field, which 
is described by the second term of Eq.(\ref{phi}), will be negligible at the horizon 
location $R_H \sim 1/ M_{\rm f}$. For the typical collider produced black holes, the singularity
will not be naked.
For the same reason, modifications of the Hawking evaporation spectrum can be neglected in
the charge range under investigation. 

Let us briefly comment on the assumption that the electromagnetic field is spherical symmetric up
to a brane width of $\sim 1/M_{\rm f}$. If the field is confined to a thinner brane, the charge contribution
to the gravitational potential will obey a different functional behaviour. It will drop slower at large distances but therefore
be less divergent at small distances. This means, if the above inequality is fulfilled it will still hold because
the singularity is even better shielded. 

Usually, the Hawking radiation for very small charged 
black holes necessarily leads to naked singularities which are hoped to be excluded
by the (unproven) cosmic censorship hypothesis.   
The reason is that once the mass of a charged black hole becomes smaller than the mass of the 
lightest charged particle - i.e. the electron or positron, respectively - it could never get rid 
of its charge by radiating it off. Then, it would either end as a naked singularity 
or as a tiny remnant of mass about the electron mass. This case, however, can not occur in the here
discussed setting as we assume the remnant mass to be close by $M_{\rm f}$ and therefore much
above the electron mass.

\section{Charged Black Hole Remnants}

Black holes are typically formed from valence quarks as those carry the largest available
momenta of the partonic system. So, the black holes formed in
a proton-proton collision will have an average charge of $\sim 4/3$. 
The black holes decay with an average multiplicity of $\approx 10 - 25$ into particles of
the SM, most of which will be charged. The details of the multiplicity
depend on the number of extra dimensions \cite{first}.
After the black holes have evaporated off enough energy to be stable at the 
remnant mass, some have accumulated a net electric charge.
According to purely statistical considerations, the probability
for being left with highly charged black hole remnants drops fast with deviation from the average. 
The largest fraction of the black holes should have charges $\pm 1$ or zero.
  
\begin{figure}[h]
\centering  \epsfig{figure=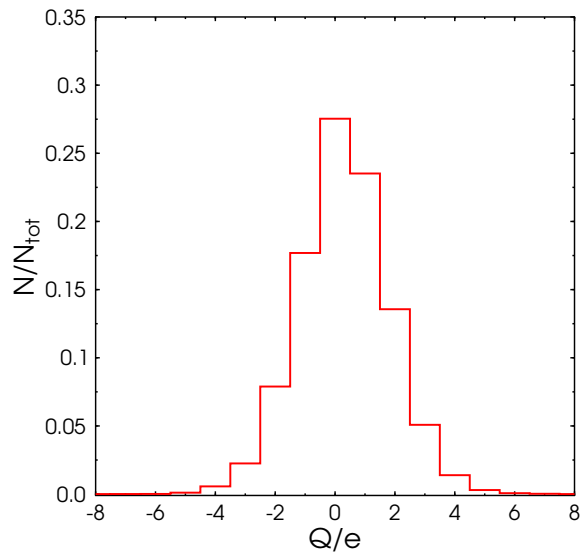,width=3.6in}
\vspace*{-0.5cm}
\caption{Distribution black hole remnant charges  in $PP$-interactions at $\sqrt{s}=14$~TeV 
calculated with the {\sc PYTHIA} event generator.\label{fig1}}
\end{figure}

For a detailed analysis, we have estimated the fraction of charged black hole remnants
with the {\sc PYTHIA} event 
generator and the CHARYBDIS program \cite{PYTHIA,CHARYBDIS}. For our purposes, we turned off the final decay of the black hole and the charge minimization.
Figure \ref{fig1} shows the results for a simulation of 
proton-proton collisions at the {\sc LHC} with an estimated center of mass
energy of $\sqrt{s}=14$~TeV.
 
We further assumed as an applicable model, worked out in \cite{first}, that the effective temperature of 
the black hole drops towards zero for a finite remnant mass $M_{\rm R}$. This mass of the remnant is a 
few $\times M_{\rm f}$ and a parameter of the model. Even though 
the temperature-mass relation is not clear from the present status of theoretical examinations, 
such a drop of the temperature can be implemented into the simulation.  
However, the details of the modified temperature  as well as the value of $M_{\rm R}$ do 
not noticeably affect the here investigated charge distribution as it results from the 
very general statistical distribution of the charge of the emitted particles.  

Therefore, independent of the underlying quantum gravitational assumption leading to the
remnant formation, we find that about $27.5$\% of the remnants carry zero electric 
charge, whereas
 we have $\approx 17.7$\% of $BH^-$ and $\approx 23.5$\% of $BH^+$.
 
The total number of produced black hole remnants depends on the total cross section for black
holes \cite{bhsonst,Chamblin:2002ad,Atlas,BHex}. Ongoing investigations on the subject reveal 
a strong dependence  on $M_{\rm f}$ and a slight
dependence on $d$ and suggest the production of $\approx 10^8$ black holes per year. 
Thus, following the above given results we predict the production of 
about $10^7$ single charged $BH^\pm$ remnants per year.

\vspace*{-1.6cm}
\hspace*{-0.3cm}

These charged black hole remnants are of special importance. Once produced, the $BH^\pm$ could be
singled out in an experiment before they are 
neutralized in a detector. The characteristically small $Q/M$ makes them easy to 
distinguish from particles of the SM. Their electromagnetic interaction further
allows it to trap and keep them 
in an electromagnetic field. 

For the specific scenario discussed here, the average momenta of the
black hole remnants are of the order of $p\sim 1$~TeV. We suggest to use a
similar approach as used for the trapping of anti-protons at
LEAR/TRAP \cite{Gabrielse:1989xw}. This means, first the remnants are
decelerated in a decelerator ring from some GeV/c down to 100~MeV/c.
Then they have to be further slowed down by electric fields to a
couple of keV.  This is slow enough to allow for a capture of the
remnants in a Penning trap with low temperature. Then positrons (or
electrons) are loaded into the trap. The positrons/electrons cool
down to the temperature of the Penning trap by the emission of
cyclotron radiation. Unfortunately, the lower cyclotron frequency of
the heavy (thus slow) remnants makes this cooling mechanism less
efficient for black hole remnants. However, they can be cooled
indirectly by Coulomb interaction with the positrons or electrons.

\vspace*{-.5cm}
\hspace*{-0.3cm}
\begin{figure}[h]
\centering  \epsfig{figure=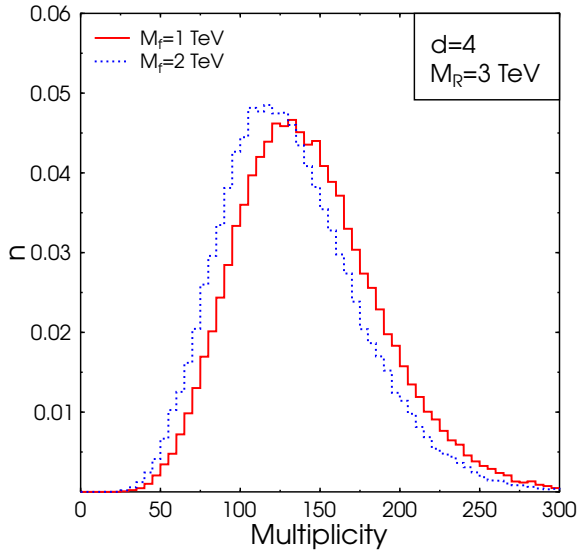,width=3.6in}
\vspace*{-0.6cm}
\caption{The distribution of events  
 over their total multiplicity after fragmentation 
for $d=4$ and $M_{\rm R}=3$~TeV (normalized to unity).  
\label{fig2}}
\end{figure}

\vspace*{-.5cm}
\hspace*{-0.3cm}
\begin{figure}[h]
\centering  \epsfig{figure=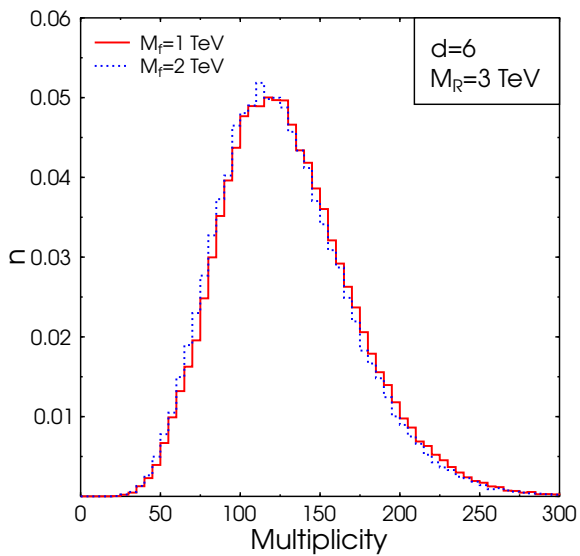,width=3.6in}
\vspace*{-0.6cm}
\caption{The distribution of events  
 over their total multiplicity after fragmentation 
for $d=6$ and $M_{\rm R}=3$~TeV (normalized to unity).  
\label{fig3}}
\end{figure}

In the case of anti-protons, the above discussed method allowed the TRAP collaboration to
store the anti-protons for many month. This time would be sufficient
to collect a huge amount of black hole remnants for study, even if
only a small percentage will have low enough energies for
deceleration.

Another approach to collect black hole remnants might be to slow down the charged remnants by
energy loss in matter - a similar approach was suggested by \cite{Arvanitaki:2005nq} to stop
gluinos. The energy loss experienced by a charged particle 
when travelling through matter can be calculated using the Bethe-Bloch equation.
>From the average momentum of the remnants, we conclude that 50\% of the remnants will 
have velocities $\beta\le 0.3$. I.e. these remnants can be decelerated in matter e.g.
in an iron block of 8~cm ($\beta=0.1$), 1.3~m ($\beta=0.2$) or 6.4~m ($\beta=0.3$) 
length. This method would allow to include even high momentum remnants into 
the trapping process.

Thus, these approaches indicates the possibility to accumulate separated
$BH^+$ and $BH^-$ over a long period of time.

In a second stage, the $BH^+$s can be merged with the $BH^-$s which increases the horizon in the process 
$BH^+ + BH^- \to BH^0$. During this process, the charge of the
forming black hole is neutralized and the mass is increased to $2 M_{\rm R}$. This will make 
a new evaporation possible which can then be analyzed in an environment clean of background from the proton-proton
collision. In particular, the characteristics of the late stages of the decay can be observed closely.

After the merged black holes have shrunk again to remnant mass, most of them will be neutral 
and escape the experimentally accessible region due to their small cross section.

\begin{figure}[h]
\centering  \epsfig{figure=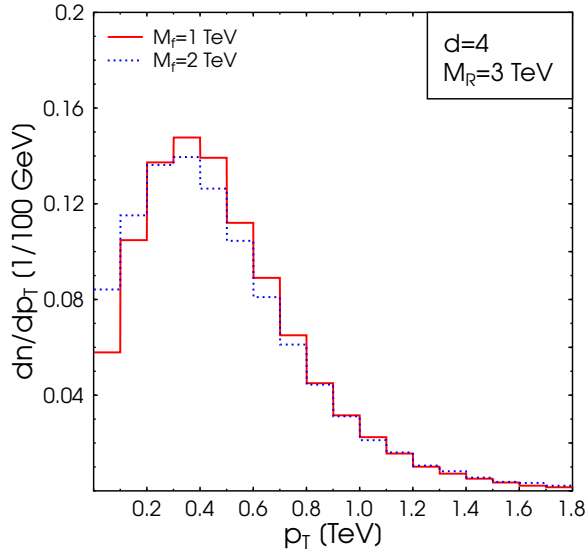,width=3.6in}
\vspace*{-0.5cm}
\caption{The (normalized) $\gamma$ - distribution over transverse momentum  
for $d=4$ with $M_{\rm R}=3$~TeV.
\label{fig4}}
\end{figure}

\begin{figure}[h]
\centering  \epsfig{figure=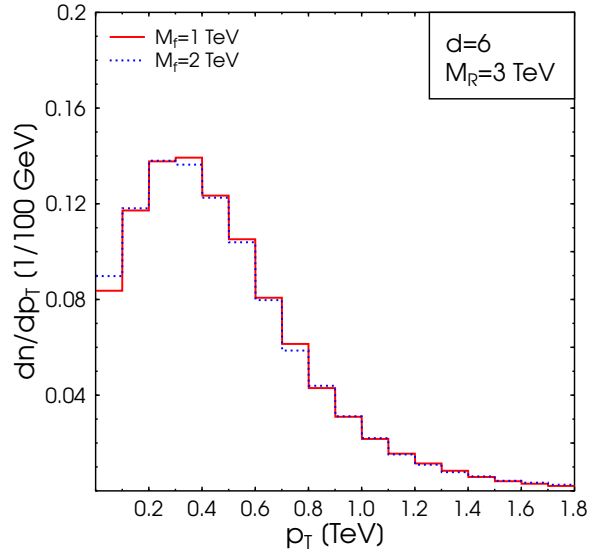,width=3.6in}
\vspace*{-0.5cm}
\caption{The (normalized) $\gamma$ - distribution over transverse momentum  
for various $d=6$ with $M_{\rm R}=3$~TeV.
\label{fig5}}
\end{figure}

Figure \ref{fig2} to \ref{fig5} show results from a simulation of such reactions for a 
sample of $10^6$ events of  $BH^+ + BH^- \to BH^0$ with CHARYBDIS modified to incorporate the
remnant production according to \cite{first}. Here, we have assumed that the black holes
have been slowed down enough to make the initial momentum negligible. The total energy
of the collision is then $2 \times M_{\rm R}$. Figure \ref{fig2} and  \ref{fig3} show the
total multiplicity of the events after fragmentation, Figures \ref{fig4} and \ref{fig5} show the $p_T$-
spectra for $\gamma$'s as an example . The spectra are free from low energetic debris 
present in $PP$ collisions arising from the presence of spectators.  
 
Even though the parameter of the model might be difficult to extract (the dependence on $d$ 
and $M_{\rm f}$ would require  initial states of varying masses) a
measurement of such
spectrum would be a very important input to examine the signatures from the $PP$ 
collision at the LHC. In such a way, the remnants would allow to extract  
the properties of the black hole's decay and remove theoretical uncertainties by
allowing to quantify them directly from experimental measurements. This would substantially
increase the precision by which the parameters of the underlying extra-dimensional
model can be determined.
 
\vspace*{-.5cm}
\hspace*{-0.3cm}

\section{Conclusion}

In the scenario of large extra dimensions, the formation of black hole
remnants in high energy collisions is possible. We have computed the
fraction of charged black hole remnants. With use of the PYTHIA event
generator we  found that about 23.5\% of the remnants carry charge $+1$ and 
17.7\% carry charge $-1$.
Due to their electromagnetic interaction, we have argued that the
black hole remnants can be trapped and then could be used to examine 
the evaporation characteristics.
In this case, the absence of background from the collision region and the 
controlled initial state would allow to obtain very clean signals. This would make 
it possible to extract information about the late stages of the evaporation process
of black holes with unprecedented  precision.

\section*{Acknowledgements}

We  thank  Horst St\"ocker  for 
helpful discussions. This work was supported by NSF PHY/0301998 and {\sc DFG}.
SH wants to thank the FIAS for kind hospitality.

\vspace*{-.5cm}
\hspace*{-0.3cm}

\end{document}